\documentstyle[11pt,twocolumn]{article}

\author{\normalsize Fedor Herbut\\
\small \it Serbian Academy of Sciences
and Arts,
 Knez Mihajlova 35, Beograd, Serbia\\
\small \it e-mail:
fedorh@infosky.net}

\def\cH{{\cal H}}
\def\cR{{\cal R}}
\def\Tr{{\rm Tr}}
\def\ket#1{\mid~\!\!\!{#1}~\!\!\rangle}
\def\bra#1{\langle~\!\!{#1}~\!\!\!\mid}

\title{\Large \bf Distinguishing quantum
measurements of observables\\
in terms of state transformers}

\date{
\begin{flushleft} \normalsize \rm
The modern framework of state
transformers, i. e., the first Kraus
representation of quantum measurement, is
introduced and related both to the known
textbook concepts and to
measurement-interaction evolution (the
second Kraus representation). In this
framework the known kinds of measurements
of ordinary (as distinct from
generalized) observables  are
distinguished by necessary and sufficient
conditions. Thus, repeatable,
nonrepeatable, and ideal measurements are
characterized both algebraically and
geometrically utilizing polar
factorization of state transformers.
\end{flushleft}}

\begin{document}

\maketitle

\normalsize
 \rm
{\bf I. INTRODUCTION}

\indent Quantum mechanics is no longer an
esoteric scientific discipline solely
about microscopic objects far beyond
human perception. Much progress has been
made in demonstrating the macroscopic
quantum behavior of various systems such
as superconductors, superfluids,
nanoscale magnets, laser-cooled trapped
ions, photons in a microwave cavity etc.
On the fundamental level quantum
mechanics is expected to underlie
classical physics.

Measurement theory is a central part of
quantum mechanics. Connection of the
quantum formalism with laboratory
practice hinges on measurement, which
gives results and thus enables one to
draw information from an observed
specimen.

The newest and, perhaps, most promising
applications of quantum mechanics -
quantum information theory and quantum
computation - also require quantum
measurement as a basic notion.$^{1,2}$

Nowadays the concepts of quantum
measurement and of the ensuing change of
state are far from those found in some
older textbooks. {\it The aim of this
article} is to give a simple presentation
of the modern framework of quantum
measurement theory, and to discuss the
basic kinds of measurements of ordinary
discrete observables in it.

In textbooks one often gives a discrete
observable (Hermitian operator) $M$ with
possibly degenerate (distinct)
eigenvalues in spectral form
$$M=\sum_im_i\sum_j\ket{ij}\bra{ij}$$
$$=\sum_im_iP_i.\eqno{(1)}$$ The range of
values of $j$ depends on the degeneracy
of $m_i$. The spectral form in terms of
the eigenprojectors $P_i$ is uniquely
determined by $M$. The one in terms of
the eigenbasis $\{\ket{ij}:\forall ij\}$
is, on account of the degeneracies,
nonunique. (Throughout, by "basis" is
meant an orthonormal (ON) set of vectors
spanning the entire space.)

Some standard texbooks$^{3,4}$ then claim
that the eigenvalues $m_i$ are the only
possible results of the measurement of
$M$, and, if the result $m_i$ is
obtained, then its probability is, of
course, positive, and an arbitrary state
vector $\ket{\psi}$ undergoes the change:
$$\ket{\psi}\quad \rightarrow \quad
\ket{\psi}_i\equiv
p_i^{-1/2}P_i\ket{\psi},\eqno{(2a)}$$
where
$$\forall i:\quad p_i\equiv \bra{\psi}P_i
\ket{\psi}\eqno{(2b)}$$ are the
probabilities for the changes (2a).

One often considers the change occurring
to the entire ensemble described by
$\ket{\psi }$ when an observable $M$ is
measured:
$$\ket{\psi}\bra{\psi}\quad \rightarrow
\quad
\sum_i(p_i\ket{\psi}_i\bra{\psi}_i)=$$
$$\sum_i(P_i\ket{\psi}\bra{\psi}P_i).
 \eqno{(2c)} $$ Finally, one has the
 completeness relation
$$\sum_iP_i=1.\eqno{(2d)}$$ (Note that if
$p_i=0$, $\ket{\psi}_i$ cannot be defined
as a unique state vector, and the
corresponding term
$(p_i\ket{\psi}_i\bra{\psi}_i)$ in (2c)
is understood to be zero.)

To avoid physically inessential
mathematical intricacies, we confine
ourselves throughout to
finite-dimensional state spaces.

Following the practical terminology used
in the textbook of Kaempffer,$^5$ we call
(2a) selective measurement, whereas we
refer to (2c) as nonselective
measurement. Note that one is dealing
with two aspects of measurement.
(Sometimes one objects to the use of the
word "measurement" for the nonselective
aspect because there is no definite
result. But this lack is only apparent
due to suppression of the role of the
measuring instrument. Nonselective
measurement is the totality of all
selective measurements, and the
individual instruments keep track of the
various definite results.)

It is often believed that formulae
(2a)-(2c) originated in the monumental
book of von Neumann.$^6$ Actually,
formulae (2a) and (2c) are due to L\"
uders.$^7$ There is a precise sense in
which formula (2c) can be interpreted as
describing minimal change of state in
measurement.$^8$ (See also the references
cited in ref. 8.)

Von Neumann discussed only measurement of
discrete observables with all eigenvalues
nondegenerate. If a discrete observable
is not of this kind, it can always be
refined into such an observable. To see
this, one has to consider the first
(nonunique) spectral form in (1) and
replace $m_i$ by distinct $m_{ij}$ in it
to obtain an observable $M'$ that has all
eigenvalues nondegenerate. $M'$ is a
maximal refinement of $M$. The operator
$M$ is a function of $M'$: $M=f(M')$
meaning that the eigenvectors of $M'$ are
eigenvectors also of $M$, and $\forall
ij:\enskip m_i=f(m_{ij})$.

Following the idea of von Neumann, one
can measure $M$ by measuring $M'$. But it
is not a minimal measurement, because
(2a) is replaced by
$$\ket{\psi} \bra{\psi}\quad \rightarrow
\enskip \sum_j(p_{ij}/p_i)
\ket{ij}\bra{ij},$$ where the
probabilities are now $$\forall ij:\quad
p_{ij}\equiv ||\bra{ij}\ket{\psi}||^2.$$
In this way one performs an excessive
measurement of $M$. (One overdoes the
measurement not only regarding the change
of state, but also regarding the
information obtained in the measurement
because one learns about an observable
$M'$ that is a refinement of $M$.)

It is perhaps interesting to point out
that the mentioned excessive measurement
of the state can always be avoided if one
chooses a suitable maximal refinement
$M'$ of $M$. But then the choice of $M'$
must depend on the state $\rho$ in which
the measurement is performed.$^9$

Owing to (2d), one can write $\ket{\psi}
\bra{\psi}= \sum_i\sum_{i'}P_i\ket{\psi}
\bra{\psi} P_{i'}$. Viewing this in
matrix form in the representation of the
eigenbasis $\{\ket{ij}:\forall ij\}$ of
$M'$, and comparing this with (2c), one
can see that all off-diagonal submatrices
are replaced by zero. Thus $\ket{\psi}$
looses coherence. By "coherence" one,
actually, means interference experiments,
in which one measures in $\ket{\psi}$
some observable incompatible with $M$ .
The same measurement performed after that
of $M$ gives quite different results.
Thus, in (2c) coherence is lost or
decoherence (with respect to the
eigenprojectors of $M$) sets in. (I
recommend as further reading ref. 10. It
gives a short and clear presentation of
the widely accepted environment-induced
decoherence theory in measurement.)

Besides the quantum system, which we
denote as subsystem $1$, there is a
measuring apparatus, subsystem $2$, with
an initial state $\ket{0}_2$, and a
certain interaction between system and
apparatus expressed as a unitary
evolution operator $U_{12}$ that takes
the initial state $(\ket{\psi}_1\otimes
\ket{0}_2)$ into a suitably correlated
{\it final composite-system state vector}
$$\ket{\Psi}^f_{12}=U_{12}
\Big(\ket{\psi}_1\otimes \ket{0}_2\Big).
\eqno{(3a)}$$

The measuring apparatus contains one more
entity: an observable
$B_2=\sum_i(b_i\ket{\chi_i}_2
\bra{\chi_i}_2)$ with all eigenvalues
nondegenerate, called the pointer
observable. (For simplicity the dimension
of the state space of the measuring
apparatus is assumed to be equal to the
number of distinct eigenvalues of $M$.)

The mentioned "suitability" of
correlations in the final state means
that if one {\it expands} the final state
vector $\ket{\Psi}_{12}^f$ in the
eigenbasis $\{\ket{\chi_i}_2:\forall i\}$
of the pointer observable, one obtains
$$\ket{\Psi}^f_{12}=\sum_i
(P_1^i\ket{\psi}_1)\otimes
\ket{\chi_i}_2.\eqno{(3b)}$$ (Note that
$P_1^i$ is the same as $P_i$. The change
of notation is due to the need to deal
with two subsystems.)

{\it Expansion} of a bipartite vector in
a {\it factor basis} is always possible
and it gives unique (generalized)
"expansion coefficients" (vectors in the
opposite factor space). One can see this
by taking a basis also in the opposite
factor space, expanding the bipartite
vector in the product basis, and,
finally, by grouping the opposite-space
vectors that go with one and the same
basis vector (in the factor space in
which the expansion is
performed).\footnote{When a composite
state vector, e. g. $\ket{\Psi}^f_{12}$,
is expanded in an arbitrary
second-subsystem basis, the second tensor
factors are orthonormal state vectors,
but the first ones need not be; they need
not even be orthogonal in general (cf (4)
e. g.). In the special case when the
composite-system state vector is expanded
in an eigenbasis of its reduced density
matrix, in our example, of $\rho_2 \equiv
\Tr_1(\ket{\Psi}^f_{12}\bra{\Psi}^f_{12})$,
where the partial trace is taken over the
first subsystem, and only in this case,
also the first factors are orthogonal
vectors. Then one can write the expansion
with positive numerical coefficients and
with ON vectors in both factors. This is
then the Schmidt expansion.$^{11,12}$ It
can also be obtained by interchanging the
role of subsystems $1$ and $2$. (Anyway,
the Schmidt expansion is biorthogonal, i.
e., orthogonal in both factors, and it is
expansion in the eigenbases of both
reduced density matrices simultaneously.)
The Schmidt expansion is often made use
of; but it will not be utilized in this
article. (Though, expansion (3b) is
obviously biorthogonal, and, as it has
just been explained, it could be easily
rewritten as a Schmidt expansion.)}

One says that one "reads the pointer
position" when one measures $B_2$
(instantaneously). If the result is
$b_i$, then, of course, $p_i>0$, and one
applies the L\" uders projection (2a)
{\it mutatis mutandis}, which gives
$$p_i^{-1/2}(1_1\otimes
\ket{\chi_i}_2\bra{\chi_i}_2)
\ket{\Psi}^f_{12}=$$ $$(p_i^{-1/2}P_i
\ket{\psi}_1) \otimes \ket{\chi_i}_2$$ in
this case. Thus one arrives at (the
literal form of) (2a).

The measurement of the pointer observable
is accompanied by collapse or
objectification$^{13}$ of the
composite-system state. These terms are
closely connected with the selective
version of decoherence, which takes place
in nonselective measurement. The physical
source of the phenomenon is a
controversial point. But it is beyond
dispute that this problem of measurement
theory arises from the incompatibility of
the unitary linear dynamics of the
composite system plus apparatus (cf (3a))
with the transition from a superposition
of pointer states (cf (3b)) to a definite
pointer state (cf (2a)).
We will not discuss the problem of collapse
in this study.\\

{\bf II. THE FIRST AND THE SECOND
REPRESENTATIONS OF KRAUS}

\indent Any choice of $U_{12}$,
$\ket{0}_2$, and a pointer observable
$B_2=\sum_ib_i\ket{i}_2\bra{i}_2$ lead,
via (3a), to some final state vector
$$\ket{\Psi}^f_{12}\equiv
\sum_i(M_i\ket{\psi}_1)\otimes \ket{i}_2,
\eqno{(4)}$$ which defines {\it linear
operators} $\{M_i:\forall i\}$ in the
state space of the system. Reading the
pointer position $b_i$ for $p_i>0$ gives
rise to the {\it selective change of
state}
$$(\ket{\psi}_1\otimes \ket{0}_2)\quad
\rightarrow \quad p_i^{-1/2}(1_1\otimes
\ket{i}_2\bra{i}_2) \ket{\Psi}^f_{12},$$
i. e., as seen from (4),
$$\ket{\psi}\quad \rightarrow \quad
p_i^{-1/2}M_i\ket{\psi}.\eqno{(5a)}$$
The probabilities are
given by $$\forall i:\enskip
p_i=\bra{\psi}
M_i^{\dagger}M_i\ket{\psi}, \eqno{(5b)}$$
where $M_i^{\dagger}$ is the adjoint of
$M_i$.  The corresponding {\it
nonselective measurement} produces the
{\it change of state}
$$\ket{\psi}\bra{\psi}\enskip \rightarrow
\enskip \Tr_2\ket{\Psi}^f_{12}
\bra{\Psi}^f_{12}$$
$$=\sum_iM_i\ket{\psi}
\bra{\psi}M_i^{\dagger}. \eqno{(5c)}$$
 Finally, one has
$$\sum_iM_i^{\dagger}M_i=1.\eqno{(5d)}$$
(This is implied by the facts that the
RHS of (5c) has trace one and that
$\ket{\psi}$ is an arbitrary state
vector.)

The operators $\{M_i:\forall i\}$ are
sometimes called {\it state transformers}
on account of the selective change of
state (5a).$^{14}$ They are also called
measurement operators.$^2$

Evidently, putting $\forall i:\enskip
M_i\equiv P_i$, relations (5a)-(5d) take
on the special form (2a)-(2d). In this
case one is dealing with the  L\" uders
state transformers. (Note that
measurement of each observable $M$ has
its own pointer basis
$\{\ket{i}_2:\forall i\}$. In the special
case at issue one must put $\forall
i:\enskip \ket{i}_2\equiv \ket{\chi}_2$).

If the quantum state is a general (mixed
or pure) one, then a density operator
$\rho$ takes the place of $\ket{\psi}$.
The reader can easily prove that
relations (5a)-(5c) generalize into
$$\mbox{if}\enskip p_i>0,\enskip
\rho \enskip \rightarrow \enskip \rho_i
\equiv p_i^{-1}M_i\rho
M_i^{\dagger},\eqno{(6a)}$$
$$\forall i:\quad p_i\equiv \Tr (\rho
M_i^{\dagger}M_i),\eqno{(6b)}$$ $$\rho
\quad \rightarrow \quad \sum_iM_i\rho M_i
^{\dagger}. \eqno{(6c)}$$ (Hint: Express
$\rho$ as a convex combination of pure
states with statistical weights as the
coefficients.)

To decide if one is now dealing with a
general measurement, one wonders about
the converse state of affairs: If
$\{N_l:\forall l\}$ is an arbitrary set
of state transformers, i. e., if they are
linear operators satisfying (5d) {\it
mutatis mutandis}, does there exist a
unitary operator $\bar U_{12}$ that will
take $(\ket{\psi}_1 \otimes \ket{0}_2)$
into a final state $\ket{\Phi}^f_{12}$
implying (5a-c) with these linear
operators {\it mutatis mutandis}?
Affirmative answer requires that the
number of values of $l$ must not exceed
the dimension of the second state space
because otherwise the following relations
(analogues of (4) and (3a) respectively)
are not consistent:
$$\ket{\Phi}^f_{12}\equiv
\sum_l(N_l\ket{\psi}_1)\otimes
\ket{l}_2\eqno{(7a)}$$
$$\equiv \bar U_{12} (\ket{\psi}_1\otimes
\ket{0}_2).\eqno{(7b)}$$ The basis
$\{\ket{l}_2:\forall l\}$ can be the same
as in (4).

{\bf Exercise 1.} Show that if
$\bra{\psi_i}\ket{\psi_j}=\delta_{i,j}$,
then also the corresponding
composite-system vectors
$\ket{\Phi}^f_{12}$ defined by relation
(7a) are orthonormal. (Hint: Utilize (5d)
{\it mutatis mutandis}.)

Thus, relation (7a) determines $\bar
U_{12}$ incompletely as an (incomplete)
isometry, i. e., a linear map taking a
subspace onto another equally dimensional
one, preserving the scalar product.

Any incomplete isometry can be extended,
though nonuniquely, into a unitary
operator in the entire space. (This is
seen by completing the pair of
orthonormal sets of vectors that
determine the partial isometry into
bases.)

Thus, the answer to the above question is
affirmative, and relation (7b) makes
sense in terms of a unitary operator
$\bar U_{12}$. The first relation defines
the composite state vector in terms of
the operators $N_l$, and the second
determines $\bar U_{12}$, though
incompletely.

Therefore, one speaks of (5a)-(5d) (or of
(6a)-(6c) and (5d)) as a {\it general
quantum measurement} or the measurement
of a general quantum observable.$^{14}$
(For details see the standard textbook of
quantum information theory$^2$, or the
enlightening review$^{15}$ and the
references therein, or ref. 16.) General
quantum measurements comprise, besides
measurements of ordinary observables,
also measurements of generalized
observables (see Definition 4 below). The
latter are closely connected with
so-called positive-operator-valued
measures (POVM). We discuss them below.

A set of state transformers on the one
hand, and the initial state of the
measuring apparatus, the unitary
interaction operator, and the pointer
observable on the other hand are two
sides of a coin. Sometimes one calls them
"instrument".$^{17}$

Putting an ordinary or generalized
measurement in the form of linear
operators $\{M_i:\forall i\}$ satisfying
(5d), i. e., expressing it in terms of
state transformers, is called the first
Kraus representation or the operator-sum
representation. The use of
composite-system unitary operators via
(3a) and (4) is called the second Kraus
representation.$^{16}$ The state
transformers are sometimes called Kraus
operators.

Instead of the considerably intricate
(but also rather general) exposition of
Kraus in ref. 16, one can read the
simpler and more modern presentation in
ref. 18 or that in ref. 1.

One should note that $\Pi_i\equiv
M_i^{\dagger}M_i$ are  positive operators
associated with the selected measurement
results. A positive operator $\Pi_i$
satisfies, by definition, the inequality
$\bra{\psi}\Pi_i\ket{\psi}\geq 0$ for
every vector $\ket{\psi}$. It is
necessarily Hermitian. It has the
important property of possessing a unique
positive (operator) square root.

On account of the positivity of the
$\Pi_i$, one speaks, in general, of {\it
positive-operator-valued measures} (POVM)
as generalizations of observables
(Hermitian operators). The latter
represent the special case of
projector-valued measures (see Definition
1 below).\footnote{In the mathematical
literature a positive-operator- (or the
special case of a projector-) valued
measure is defined on the set of all
Borel sets (generalizations of intervals)
of the real line, and then it is
equivalent to a general (or ordinary)
observable. In the case of discrete
observables, to which we are confined in
this article, besides $\{\Pi_i:\forall
i\}$ (or $\{P_i:\forall i\}$), it is also
indispensable to specify the (real)
eigenvalues $\{m_i:\forall i\}$.}

{\bf Lemma}: Whenever a POVM $\{\Pi_i:
\forall i\}$ is given, there exist linear
operators $\{M_i:\forall i\}$ such that
$\forall i:\enskip
M_i^{\dagger}M_i=\Pi_i$.

{\it Proof}: Let $\{U_i:\forall i\}$ be
arbitrary unitary operators, and let us
define $\forall i:\enskip M_i\equiv
U_i(\Pi_i)^{1/2}$. Then
$M_i^{\dagger}M_i=(\Pi_i)^{1/2}U_i^{-1}U_i
(\Pi_i)^{1/2}=\Pi_i$. (Naturally, one can
take all $U_i$ equal to $1$.) \hfill
$\Box$

{\bf Remark 1}: Thus, any POVM defines a
whole family of state transformers, all
of which reproduce the same POVM via
$\Pi_i\equiv M_i^{\dagger}M_i$. The
probabilities, which equal
$$\forall i:\quad p_i\equiv
\bra{\psi}\Pi_i\ket{\psi},\eqno{(8)}$$
depend only on the positive operators
$\Pi_i$ associated with the individual
results. But the changes of state, e. g.,
the selective ones, depend on the linear
operators $M_i$ (cf (5a)), and these are
associated with the $\Pi_i$ in a
nonunique way. One should keep in mind
that it is actually the interaction
between system and measuring apparatus
(plus the pointer basis) that selects the
$M_i$, because it determines the unitary
evolution operator $U_{12}$, and this, in
turn, determines the linear operators
$M_i$ via (3a) and (4).

General measurement or unitary evolution,
i. e., any quantum mechanical change of
state, is expressible in the form of
action of a trace preserving completely
positive superoperator on the density
operator of the system. (A positive
superoperator, by definition, preserves
positivity of the operator on which it
acts. Complete positivity means that even
when the positive superoperator is tensor
multiplied by the identity superoperator,
the resulting composite-system
superoperator is also positive.) Every
such superoperator is amenable to both
the first and the second Kraus
representations.

It is widely accepted that POVM and the
two Kraus representations, i. e., the
language of state transformers and
composite-system unitary operators, is
the modern framework of quantum
measurement
theory.\\

{\bf III. THE SPECIAL CASE OF ORDINARY
OBSERVABLES}

\indent Let the measurement of a general
observable, i. e., that of an ordinary or
a generalized one, be given in terms of
state transformers $\{M_{i'}:\forall
i'\}$, which imply the POVM
$\{\Pi_{i'}\equiv M_{i'}^{\dagger}M_{i'}:
\forall i'\}$. On the other hand, let a
resolution of the identity
$\sum_{i'}P_{i'}=1$ enumerated by the
same index be given. (The $P_{i'}$ are
orthogonal projectors; physically:
disjoint events, alternatives in suitable
measurement, altogether making up the
certain event) .

{\bf Definition 1}: Let the POVM for a
given index value $i$ equal $P_i$. Then
the state transformers $\{M_{i'}:\forall
i'\}$ imply the selective measurement of
an ordinary discrete observable
(Hermitian operator)
$M=\sum_{i'}m_{i'}P_{i'}$ (cf (1)), in
particular, of the eigenevent $P_{i}$ .
(By "eigenevent" is meant the physical
interpretation of the eigenprojector.) If
the POVM reduces to a projector-valued
measure for all values of $i'$, then the
state transformers represent a
nonselective measurement of $M$.

If one deals with the selective
measurement of an observable $M$ with the
eigenprojector $P_i$, then the
probability relation (6b) takes on the
familiar form:
$$p_i=\Tr (P_i\rho ).
\eqno{(9)}$$

One should note that one is dealing with
a class of discrete Hermitian operators
defined by a common resolution of the
identity $\sum_iP_i=1$. An arbitrary
element $M$ of the class is obtained by
associating  a distinct real number $m_i$
with each value of $i$ (cf (1)).

{\bf Definition 2}: If the measurement of
$m_i$ of $M$ has the property that its
repetition necessarily gives the same
result, then we have {\it repeatable
measurement}. More precisely, if the
measurement of the value $m_i$ of an
observable $M$ given by (1) is expressed
by the state transformer $M_i$ from a set
of state transformers $\{M_{i'}:\forall
i'\}$, and the probability of the same
value $m_i$ of $M$ in the transformed
state $M_i\rho M_i^{\dagger}/p_i$ is $1$
for every state $\rho$, then the
(selective) measurement is called
repeatable. Otherwise, it is called {\it
nonrepeatable}. If the selective
measurements are repeatable for all
values $m_i$, then the nonselective
measurement is said to be repeatable.

Synonyms for "repeatable measurement" are
"predictive", and "first-kind
measurement". Nonrepeatable measurements
are also said to be retrodictive or
retrospective or of the second kind.
(These concepts are not equivalent in the
case of generalized observables.$^{14}$)

The most usual kind of measurement in the
laboratory are the nonrepeatable ones.
Various kinds of detectors (ref. 19,
chapter 5) and the Wilson cloud chamber
(ref. 19, p. 154), e. g., measure the
instantaneous position of the particle.
But after the measurement, the state of
the particle is changed in a,
practically, unknown way. The detailed
form of the evolution operator $U_{12}$
(cf (3a)) is, of course, not known.
Hence, the state transformers (cf (4))
cannot be evaluated.

Search for laboratory realizations of
repeatable measurement goes under the
term nondemolition measurement (the value
of the measured observable is not
demolished).$^{20,21}$

{\bf Definition 3}: Let a set of state
transformers $\{M_i:\forall i\}$ be given
such that (i) they define nonselective
measurement of an ordinary discrete
observable $M$ (given by (1)), (ii) the
measurement is repeatable, and (iii) it
is a minimal-disturbance measurement in
the sense that the sharp value of every
observable compatible with the measured
one is preserved in the nonselective
measurement. (By a "sharp" value one
means one with zero dispersion or zero
variance, i. e., an eigenvalue of the
corresponding operator.) Then one has
{\it ideal measurement} of $M$.

{\bf Theorem 1}: (i) Let a general
measurement $\{M_i:\forall i\}$ be the
measurement of the value $m_i$ of an {\it
ordinary discrete observable} given by
(1), i. e., let
$$M_i^{\dagger} M_i=P_i.\eqno{(10)}$$ be
valid. (Operator form of (6b) and (9)
together.) Then one has
$$M_i=M_iP_i.\eqno{(11)}$$

(ii) {\it If and only if} besides (10)
also
$$M_i=P_iM_i\eqno{(12)}$$ is valid, the
selective measurement is {\it
repeatable}. Otherwise, it is
nonrepeatable.

(iii) The nonselective repeatable
measurement of $M$ is {\it ideal if and
only if}
$$\forall i:\enskip M_i=P_i,\eqno{(13)}$$
i. e., if the $M_i$ are the L\" uders
state transformers.

{\it Proof}: (i) If (10) is valid, then
the square norm of the vector
$M_i(P_i^{\perp}\ket{\psi }$, where
"$\perp$" denotes the orthocomplement, is
zero: $\bra{\psi
}P_i^{\perp}M_i^{\dagger}M_iP_i^{\perp}
\ket{\psi }=\bra{\psi
}P_i^{\perp}P_iP_i^{\perp} \ket{\psi
}=0$. Hence, also the vector itself is
zero, and $M_i\ket{\psi
}=M_i(P_i+P_i^{\perp})\ket{\psi }=M_iP_i
\ket{\psi }$. Since this is valid for an
arbitrary vector $\ket{\psi }$, the
claimed relation (11) follows.

(ii) As it is well known, an event
(projector) $P$ is predicted with
certainty in a state $\ket{\psi }$ if and
only if $P\ket{\psi }=\ket{\psi }$.
Hence, according to Definition 2, the
measurement at issue is repeatable if and
only if $P_iM_i\ket{\psi }= M_i\ket{\psi
}$ (cf (5a)). Since the state vector
$\ket{\psi }$ is arbitrary, the claimed
criterion (12) ensues.

(iii) It has been proved in the detailed
modern measurement theory given in ref.
13 (chapter III, subsection 3.7) that the
state transformers of ideal measurement
are those given by (13). \hfill $\Box$

{\bf Corollary}: If a state $\rho$ has a
sharp value $m_i$ of an ordinary
observable $M$ (given by (1)), then ideal
nonselective measurement of $M$ does not
change this state at all.

{\it Proof}: Let a pure state $\ket{\psi
}$ have the sharp value $m_i$ of $M$.
Then, $M\ket{\psi }=m_i\ket{\psi }$, i.
e., $\ket{\psi }\bra{\psi }$ is an
eigenprojector of $M$, and hence it
commutes with $M$. Its sharp value $1$
has to be preserved in the corresponding
state transformation (2c) (cf Theorem
1(iii)) on account of the definition of
ideal measurement. Since $\ket{\psi }$ is
the only state vector with this property,
one has $\sum_iP_i\ket{\psi}
\bra{\psi}P_i=\ket{\psi }\bra{\psi}$.

Let $\rho$ be a mixed state with the
sharp value $m_i$ of $M$. Let us,
further, perform ideal nonselective
measurement of $M$ on $\rho$. According
to (2c), this changes the state into
$\sum_i(P_i\rho P_i)$. Writing $\rho$ in
spectral form in terms of eigenprojectors
$\rho =\sum_k(r_k \ket{k}\bra{k})$, the
transformed state is
$\sum_k\{r_k[\sum_i(P_i
\ket{k}\bra{k}P_i)]\}$. If $\rho$ has the
claimed sharp value, then necessarily the
same is true for each pure state
$\ket{k}$ (see Appendix A). Hence, as
shown in the proof in the preceding
passage, they do not change. Therefore,
neither does $\rho$.\hfill $\Box$

On account of the no-change property
expressed in the Corollary ideal
measurement is hard to find in the
laboratory. In practice the interactions
that cause a change in the measuring
apparatus (show a result) do change also
the state of the quantum system. It is
like a kind of action and reaction (like
in the second law of Newton in classical
mechanics). Theoretically there is no
problem in writing down $U_{12}$
violating this principle (if we may call
it so). (See ref. 1 for a modern
presentation of the von Neumann
procedure$^6$ of defining such a
$U_{12}$.)

As it was seen in Definition 3, the
concept of ideal measurement was
restricted to the nonselective case. But,
formally, if a state transformer is the
L\" uders one, we can speak of selective
ideal measurement. For this there exists
a laboratory realization: negative
measurement. To give a simple example,
let the Stern-Gerlach apparatus (cf ref.
3, p. 388), measuring the spin of a
spin-one-half particle, be so adjusted
that one knows when the particle enters
it. Further, let the lower half of the
screen be a detector, and let the upper
half be removed (letting the particle out
of the apparatus without interacting with
it). Then, if the particle is not
detected when it is expected in the lower
half space, it must be in the upper one,
and therefore it must have spin up. This
device is sometimes called the
Stern-Gerlach preparator because it
prepares the particle in the spin-up
state.

One may wonder why most textbooks confine
their presentation to just one kind of
measurement, the ideal one, often
disregarding the fact that it is almost
impossible to realize it in practice. The
answer, of course, is that until the last
two decades (cf ref. 16) the only known
state transformers were the L\" uders
ones. So, there was no framework for a
more general theory of measurement.

{\bf Definition 4}: If (10) does not
hold, i. e., if $\Pi_i\enskip (\equiv
M_i^{\dagger}M_i)$ is a positive operator
more general than a projector, then one
speaks of a (selective) measurement of a
{\it generalized observable}$^{14}$ or,
shortly, of a generalized measurement.

If one wonders why one needs generalized
observables, which are mentioned but not
actually discussed in this article, then
ref. 14 may be a useful source of information.\\

{\bf IV. POLAR FACTORIZATION OF STATE
TRANSFORMERS
}

\indent  Every linear operator $A$ can be
written as a product of a {\it unitary
operator} $U$ and a {\it positive} one
$H$: $A=UH$. This is called {\it the
polar factorization of the operator} (for
proof and details see Appendix B).

{\bf Theorem 2}: Let $\{U_iH_i:\forall
i\}$ be a general measurement with the
state transformers written in terms of
their polar factors.

(i) The result corresponding to $i$ has
the meaning of {\it ordinary measurement}
if and only if there exists an event
(projector) $P_i$ so that
$$H_i=P_i.\eqno{(14)}$$

(ii) If the result corresponding to $i$
has the meaning of ordinary measurement,
it is {\it repeatable} if and only if the
unitary polar factor $U_i$ maps the range
$\cR(P_i)\enskip \Big(\equiv
\{P_i\ket{\psi}:\forall
\ket{\psi}\}\Big)$
 into itself.

(iii) A nonselective repeatable
measurement of $M$ (given by (1)) is {\it
ideal} if and only if $$\forall i:\quad
U_i=1.\eqno{(15)}$$

{\it Proof}: (i) Since
$H_i^2=M_i^{\dagger}M_i$ is an identity,
the first claim is obvious in view of
(10). (ii) The sufficiency of the claimed
condition for repeatability is also
obvious because, if valid, one can write
$U_iP_i=(P_iU_i)P_i$, i. e., relation
(12) follows. Conversely, if $\ket{\psi
}\in \cR(P_i)$, then $U_i\ket{\psi }=U_i(
P_i\ket{\psi })$. Further, (14) and (12)
imply $U_i\ket{\psi}=
(P_iU_iP_i)\ket{\psi}=P_iU_i\ket{\psi}$.
This amounts to $(U_i\ket{\psi })\in
\cR(P_i)$ as claimed. (iii) The
equivalence of (13) and (15) is
obvious.\hfill $\Box$

The reader is encouraged to show that
commutation
$$[U_i,P_i]=0 \eqno{(16)}$$ is necessary
and sufficient for repeatability of
selective measurement of an observable
$M$ given by (1). (Hint: Utilize the
geometrical equivalent of (16), i. e.,
express this relation in terms of the
eigensubspaces of $P_i$.)

{\bf Remark 2}: Condition (11) is
necessary but not sufficient for ordinary
selective measurement, i. e., for (10).
To see this, let $Q_i$ be the range
projector of $H_i$, the Hermitian polar
factor of an arbitrary state transformer
$M_i$. Let, further, $E_i$ be any
projector such that $Q_iE_i=Q_i$
(geometrically: $\cR(Q_i)\subseteq
\cR(E_i)$, where, e. g., $\cR(Q_i)$ is
the range or support
$\{Q_i\ket{\psi}:\forall \enskip
\ket{\psi}\}$). Then $M_iE_i=(U_iH_i)E_i=
U_i(H_iQ_i)E_i=U_iH_iQ_i=M_i$, i. e.,
$Q_i$, and infinitely many other
projectors $E_i$ satisfy (11) for an
arbitrary state transformer $M_i$ if
$H_i$ is singular, i. e. if it has a
nontrivial null space.

In some cases state transformers may be
more easily defined when written in terms
of polar factors. To illustrate this
claim, a simple example is given (in the
form of a problem) in Appendix C.

In conclusion to this article, two points
should be emphasized. Firstly, contrary
to the impression that one may gain from
most textbooks, measurement is not
necessarily ideal; there are other, more
complex, measurements of ordinary quantum
observables (not to mention generalized
measurements). The latter are more
important from the laboratory point of
view. Secondly, the modern framework of
state transformers enables one to
distinguish the various kinds of
measurements in a simple way.\\

{\bf APPENDIX A}

\indent {\it Proof} of the general
statement that if a mixed state $\rho$
has a sharp value $m_i$ of an observable
$M$ (cf (1)) and a decomposition of this
state into pure ones $\rho
=\sum_kw_k\ket{k}\bra{k}$ is given
($\forall k:\enskip 0<w_k\leq 1,\quad
\sum_kw_k=1$), then also each pure state
$\ket{k}$ has the sharp value of the same
observable.

The state $\rho$ has the sharp value
$m_i$ of $M=\sum_{i'}m_{i'}P_{i'}$ (cf
(1)) if $\Tr \rho P_i=1$. Substituting
here the above decomposition of $\rho$
into pure states and taking into account
that $\sum_kw_k=1$, one obtains
$\sum_kw_k(1-\bra{k}P_i\ket{k})=0$. Then,
in view of $\forall k:\enskip
\bra{k}P_i\ket{k}=||P_i\ket{k}||^2\leq
1$, valid for every projector, one ends
up with $\forall k:\enskip
\bra{k}P_i\ket{k} =1$. Thus, the value
$m_i$ is sharp also in every pure state
$\ket{k}$.\hfill $\Box$\\

{\bf APPENDIX B}

{\bf The unique and the nonunique polar
factorizations}

Let $A$ be an arbitrary linear operator
in a finite-dimensional linear unitary
space $\cH$. Clearly, $A^{\dagger}A$ is a
positive operator. Let $Q$ be its range
projector (taking $\cH$ onto the range
$\cR(A^{\dagger}A)$).

{\bf Exercise 2}: Show that one can write
$A=AQ$. (Hint: Since $Q+Q^{\perp}=1$
($Q^{\perp}$ being the orthocomplementary
projector), show the equivalent relation
$AQ^{\perp}=0$, i. e., that $A^{\dagger}
A\ket{\psi }=0$ implies $A\ket{\psi }=0$.
To this purpose utilize the positive
definiteness of the norm in $\cH$.)

Utilizing (1) {\it mutatis mutandis}, one
has $A^{\dagger}A=\sum_im_iQ_i=\sum_im_i
\sum_j\ket{ij}\bra{ij}$, a spectral form
in terms of an eigenbasis
$\{\ket{ij}:\forall ij\}$ in
$\cR(A^{\dagger} A)$, distinct and
positive eigenvalues $m_i$ and
eigenprojectors $Q_i$, which are uniquely
determined by the operator
$A^{\dagger}A$. (Note that the ranges of
the values of $j$ depend, in general, on
the value of $i$.)

{\bf Exercise 3}: Show that one can write
$$A=[A(\sum_im_i^{-1/2}Q_i)]\times$$
$$[\sum_im_i^{1/2}Q_i]\equiv
\tilde UH\eqno{(A.1)}$$
in order to define $\tilde U$ and $H$.

{\bf Exercise 4}: Show that the second
factor on the RHS of (A.1) is
$H=(A^{\dagger}A)^{1/2}$, and the first
factor $\tilde U$ is a partial isometry,
i. e., a linear map in $\cH$ that takes a
subspace of $\cH$, in particular, $\cR(
A^{\dagger}A)$, onto another preserving
the value of the scalar product, and,
besides, it takes into zero the subspace
orthocomplementary to $\cR(
A^{\dagger}A)$. (Hint: Show that,
denoting by $(\dots , \dots )$ the scalar
product, one has
$(A\ket{ij},A\ket{i'j'})=m_i^{1/2}
m_{i'}^{1/2}\delta_{i,i'}\delta_{j,j'}$,
and then $(\tilde U\ket{ij},\tilde U
\ket{i'j'})=\delta_{i,i'}\delta_{j,j'}
=(\ket{ij},\ket{i'j'}$.)

The polar factors in (A.1) are evidently
uniquely determined by $A$.

{\bf Exercise 5}: Show that one can
extend $\tilde U$ into a unitary operator
$U$ in $\cH$, and that $U$ is nonuniquely
determined by $A$ unless the latter is
nonsingular. (Hint: Utilize an arbitrary
incomplete isometry mapping $\cR(
A^{\dagger}A)^{\perp}$ onto the subspace
$[\tilde U\cR(A^{\dagger}A)]^{\perp}$
preserving the scalar product in $\cR(
A^{\dagger}A)^{\perp}$ and taking
$\cR(A^{\dagger}A)$ into zero.) Further,
show that one can rewrite (A.1) in the
form
$$A=UH. \eqno{(A.2)}$$

Polar factorization (A.2), though in
general nonunique, can be more practical
than (A.1), because a unitary operator is
a concept that is simpler and more
familiar
than the notion of a partial isometry.\\

{\bf APPENDIX C}

{\bf An example of two spin-one-half
particles}

We consider the tensor product of two
(two-dimensional) spin-one-half state
spaces $\cH_1\otimes \cH_2$. We utilize
the basis $\{\ket{n}_1\otimes
\ket{n'}_2:n,n'=\uparrow ,\downarrow\}$,
which consists of tensor products of
single-particle state vectors that are
spin-up or spin-down along the z-axis
(the standard basis). Denoting by
$\sigma$ the Pauli matrix (in standard
representation), we want to specify a
simple observable:
$$M\equiv \sigma_1\otimes
1_2=P_{\uparrow}-P_{\downarrow},$$ where
both for $n=\uparrow$ and $n=\downarrow$
$$P_n\equiv \ket{n}_1\bra{n}_1\otimes
1_2.$$

Show that: (i) If one defines the state
transformers as $M_n\equiv P_n,\enskip
n=\uparrow ,\downarrow$, then one has
ideal measurement of $M$. (ii) If one
defines $M_n\equiv (1_1\otimes
U_2(n))P_n,\enskip n=\uparrow
,\downarrow$, where $U_2(n)$ is an
arbitrary unitary operator in $\cH_2$,
then one deals with repeatable selective
measurement of $M$. If all $M_n$ are
defined in this way, and the $U_2(n)$ are
chosen separately for each value of $n$,
one has nonselective  repeatable
measurement. (iii) Finally, if one
defines $M_n\equiv U_{12}(n)P_n,\enskip
n=\uparrow ,\downarrow$, where
$U_{12}(n)$ is an arbitrary unitary
operator in the composite space but such
that it does not act as the identity
operator in $\cH_1$, then one has
nonrepeatable selective measurement. If
all $M_n$ are defined in this way, and
the operators $U_{12}(n)$ are chosen
separately for each $n$, we have
nonselective nonrepeatable measurement
of $M$.\\

\scriptsize

\setlength{\parindent}{1ex}
 $^1$J. Preskill, {\it Quantum Computing
 Lectures}
 (http://www.theory.caltech.edu/people/preskill/ph229/)
 1997, chapter 3.

\setlength{\parindent}{1ex} $^2$M. A.
Nielsen and I. L. Chuang, {\it Quantum
Computation and Quantum Information}
(Cambridge Univ. Press, Cambridge UK)
2000, subsection 2.2 .

\setlength{\parindent}{1ex} $^3$C.
Cohen-Tannoudji, B. Diu and F. Lalo\" e,
{\it Quantum Mechanics} Volume I (John
Wiley and Sons, New York) 1977, chapter
III, section C.

\setlength{\parindent}{1ex}
 $^4$A. Messiah,
{\it Quantum Mechanics} (North Holland,
Amsterdam) 1961, vol. I, chapter VIII,
subsection I.2.

\setlength{\parindent}{1ex}
 $^5$F. A. Kaempffer,
{\it Concepts in Quantum Mechanics}
(Academic Press, New York) 1965, chapters
5 and 6.

\setlength{\parindent}{1ex}
 $^6$J. von Neumann {\it Mathematical
 Foundations of Quantum Mechanics}
 (Princeton University Press, Princeton)
 1955.

\setlength{\parindent}{1ex} $^7$G.
L\"{u}ders, "On the change of state in
the process of measurement" (in German),
Ann. der Physik, 8, 322-328 (1951).

\setlength{\parindent}{1ex}
 $^8$F. Herbut, "Derivation of the change
 of state in measurement from the concept
 of minimal measurement", Ann. Phys., 55,
 271-300 (1969).

\setlength{\parindent}{1ex} $^9$F.
Herbut, "Minimal-disturbance measurement
as a specification in von Neumann's
quantal theory of measurement", Intern.
J. Theor. Phys., 11, 193-204 (1974).

\setlength{\parindent}{1ex} $^{10}$W. H.
Zurek, "Decoherence and the transition
from quantum to classical", Physics Today
44, October, 36-44 (1991).

\setlength{\parindent}{1ex} $^{11}$A.
Peres, {\it Quantum Theory: Concepts and
Methods} (Kluwer Ac. Publ., Dordrecht)
1993, chapter 5, section 3.

\setlength{\parindent}{1ex} $^{12}$F.
Herbut and M. Vuji\v{c}i\'{c}, "Distant
measurement", Ann. Phys. (N. Y.), 96,
382-405 (1976).

\setlength{\parindent}{1ex} $^{13}$P.
Busch, P. J. Lahti, and T. Mittelstaedt,
{\it The Quantum Theory of Measurement},
Lecture Notes in Physics, M2 (Springer
Verlag, Berlin) 1991.

P. Mittelstaedt, {\it The Interpretation
of Quantum Mechanics and the Measurement
Process} (Cambridge University Press,
Cambridge) 1998.

\setlength{\parindent}{1ex} $^{14}$P.
Busch, M. Grabowski and P. J. Lahti,
"Repeatable measurements in quantum
theory: their role and feasibility",
Found. Phys. 25 1239-1266 (1995).

\setlength{\parindent}{1ex} $^{15}$V.
Vedral, "The role of relative entropy in
quantum information theory", Rev. Mod.
Phys., 74, 197-234 (2002).

\setlength{\parindent}{1ex} $^{16}$K.
Kraus, {\it States, Effects, and
Operations} (Springer-Verlag, Berlin)
1983.

\setlength{\parindent}{1ex} $^{17}$E. B.
Davies, {\it Quantum Theory of Open
Systems} (Academic Press, New York) 1976,
chapter 4, section 1.

\setlength{\parindent}{1ex} $^{18}$B.
Schumacher, "Sending entanglement through
noisy quantum channels", Phys. Rev. A 54,
2614-2628 (1996).

\setlength{\parindent}{1ex} $^{19}$A. E.
S. Green, {\it Nuclear Physics}
(McGraw-Hill, New York) 1955.

\setlength{\parindent}{1ex} $^{20}$V. B.
Braginsky, Y. I. Vorontsov, and K. S.
Thorne, "Quantum nondemolition
measurements", Science 209, 547-557
(1980). Reprinted in J. A. Wheeler and W.
H. Zurek editors, {\it Quantum Theory and
Measurement} (Princeton University Press,
Princeton) 1983, p. 749.

\setlength{\parindent}{1ex} $^{21}$F.
Harrison, "Measure for measure in quantum
theory", Physics World March, 24-25
(1996).

\end{document}